\documentclass[11pt]{article}
\pdfoutput=1
\usepackage[centertags]{amsmath}
\usepackage[square, comma, sort&compress,numbers]{natbib}
\usepackage{array,multirow}
\numberwithin{equation}{section}
\usepackage{amssymb}
\usepackage{graphicx}
\usepackage{color}
\usepackage{setspace}
 
\usepackage{epsfig}

\def\Or[#1]{{\text{O}}\left({#1}\right)}
\def\dotl[#1,#2]{\left\langle #1, #2 \right\rangle}
\def\dotlb[#1,#2]{[ #1, #2 ]}
\def\dotp[#1,#2]{(#1) \cdot (#2)}
\def\aff[#1,#2]{\hat{#1}(#2)}
\def\n4sym{{\cal N}=4 SYM}
\def\>{\rangle}
\def\<{\langle}
\def\weight[#1,#2,#3]{\{(#1),#2,#3\}}
\def\ads[#1]{$\text{AdS}_{#1}$}

\newcommand{\ba}{\begin{eqnarray}}
\newcommand{\ea}{\end{eqnarray}}
\newcommand{\be}{\begin{eqnarray}}
\newcommand{\ee}{\end{eqnarray}}
\newcommand{\bq}{\begin{equation}}
\newcommand{\eq}{\end{equation}}

\newcommand{\CL}{{\cal L}}

\newcommand{\CN}{{\cal N}}
\newcommand{\CO}{{\cal O}}
\newcommand{\CP}{{\cal P}}

\newcommand{\CV}{{\cal V}}
\newcommand{\nn}{\nonumber}

\newcommand\oo\infty
\newcommand\s\sigma
\newcommand\de\delta
\newcommand\De\Delta

\newcommand\f\phi
\newcommand\g\gamma
\newcommand\x\times

\newcommand{\fr}{\frac}
\newcommand{\comm}[2]{[#1,#2]}

\usepackage{jheppub}

\makeatletter
\def\@fpheader{\vspace{-.1cm}}
\makeatother

\title{Virasoro conformal blocks and thermality from classical background fields}
\title{Virasoro conformal blocks and thermality from classical background fields}
\author[a,b]{A.\ Liam Fitzpatrick,}
\author[c]{Jared Kaplan,}
\author[d]{and Matthew T.\ Walters}
\affiliation[a]{Stanford Institute for Theoretical Physics, Stanford University, \\
Via Pueblo, Stanford, CA 94305, U.S.A.}
\affiliation[b]{SLAC National Accelerator Laboratory, \\
Sand Hill Road, Menlo Park, CA 94025, U.S.A.}
\affiliation[c]{Department of Physics and Astronomy, Johns Hopkins University, \\
Charles Street, Baltimore, MD 21218, U.S.A.}
\affiliation[d]{Department of Physics, Boston University, \\
Commonwealth Avenue, Boston, MA 02215, U.S.A.}
\emailAdd{fitzpatr@stanford.edu}
\emailAdd{jaredk@pha.jhu.edu}
\emailAdd{mtwalter@bu.edu}
\abstract{
We show that in 2d CFTs at large central charge, the coupling of the stress tensor to heavy operators can be re-absorbed by placing the CFT in a non-trivial background metric.  This leads to a more precise computation of the Virasoro conformal blocks between heavy and light operators, which are shown to be equivalent to global conformal blocks evaluated in the new background.  We also generalize to the case where the operators carry $U(1)$ charges.   The refined Virasoro blocks can be used as the seed for a new Virasoro block recursion relation expanded in the heavy-light limit.   We comment on the implications of our results for the universality of black hole thermality in AdS$_3$, or equivalently, the eigenstate thermalization hypothesis for CFT$_2$ at large central charge. 
}
\arxivnumber{1501.05315}

\begin{document}

\maketitle
\flushbottom

\section{Introduction}

In gravitational theories the localized high-energy states are black holes, characterized by a universal Hawking temperature.  More generally, in any sufficiently complex theory, it has been hypothesized \cite{ETH, ETH2} that high-energy microstates behave like a thermal background.   We will derive more precise versions of these statements in 2d Conformal Field Theories (CFT$_2$), which are believed \cite{Maldacena:1997re, Witten, GKP} to also describe all consistent theories of quantum gravity in 3d Anti-de Sitter (AdS$_3$) spacetime.

States in CFT$_2$ can be decomposed into irreducible representations of the Virasoro symmetry algebra.  Correlation functions can be written as a sum over the exchange of these irreducible representations, which are individually called conformal partial waves or Virasoro conformal blocks.  The conformal blocks are atomic ingredients in the CFT bootstrap \cite{FerraraOriginalBootstrap1, PolyakovOriginalBootstrap2, Rattazzi:2008pe}, and are particularly crucial for obtaining analytic constraints on CFTs \cite{Heemskerk:2009pn, Heemskerk:2010ty, ElShowk:2011ag, Liendo:2012hy, AdSfromCFT, Fitzpatrick:2012yx, KomargodskiZhiboedov, Fitzpatrick:2013sya, Alday:2013cwa,  Alday:2014qfa, Beem:2013qxa, Beem:2014zpa, Alday:2014tsa, Chester:2014fya, Chester:2014mea, Jackson:2014nla, Fitzpatrick:2014vua}.  Via the AdS/CFT correspondence, one can interpret the Virasoro blocks as the exchange of a sum of AdS wavefunctions.  The states in the Virasoro block correspond with the wavefunctions of some primary object in AdS$_3$, plus any number of AdS$_3$ gravitons.  In other words, the Virasoro blocks capture quantum gravitational effects in AdS$_3$, providing a way to precisely sum up all graviton exchanges.  The Virasoro blocks have also been used to study entanglement entropy \cite{RT1, RT2, Headrick,HartmanLargeC,Faulkner:2013yia, TakayanagiExcitedStates, DongGravityRenyi, Chen:2013kpa, Lewkowycz:2013nqa, KrausBlocks} and scrambling \cite{Roberts:2014ifa} in AdS/CFT, both in the vacuum and in the background of a heavy pure state \cite{HartmanExcitedStates}.  There has also been significant recent progress \cite{Hellerman:2009bu, Hartman:2014oaa, Keller:2014xba} in using modular invariance to understand the spectrum of CFT$_2$ at large central charge.

\begin{figure}[t!]
\begin{center}
\includegraphics[width=0.35\textwidth]{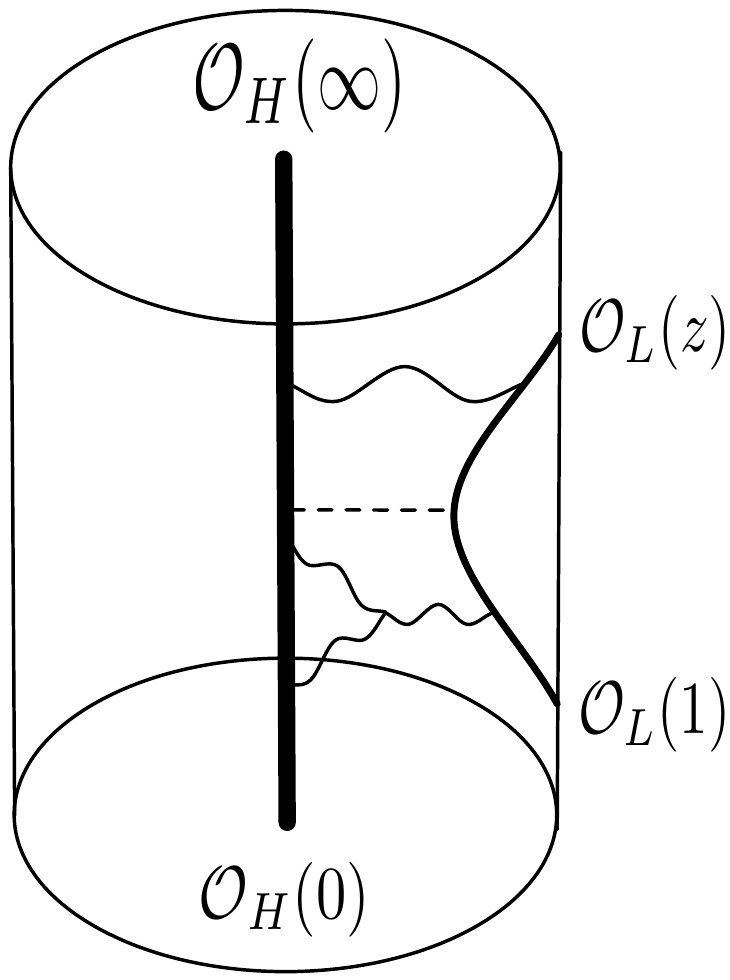}
\caption{ This figure suggests an AdS picture for the heavy-light CFT correlator -- a light probe interacting with a heavy particle or black hole.  We will show that the Virasoro blocks can be computed in this limit by placing the CFT$_2$ in an appropriately chosen 2d metric, which is related to the flat metric by a conformal transformation.
}
 \label{fig:HeavyLightCorrelatorOnAdSCylinder} 
\end{center}
\end{figure}

We will present a  new method for computing the Virasoro conformal blocks in the limit of large central charge
\be
c = \frac{3}{2 G} \gg 1,
\ee
which is often called the semi-classical limit, and for correlation functions
\be
 \< \CO_{H_1}(0) \CO_{H_2}(\infty) \CO_{L_1}(1) \CO_{L_2}(z) \> \ \ \ &\mathrm{with}& \ \ \ \frac{h_H}{c}, h_L, \delta_H, \delta_L \textrm{ fixed }, \nn\\
h_{H,L} \equiv \frac{h_{H_1,L_1}+ h_{H_2,L_2}}{2}, &\qquad& \delta_{H,L} \equiv \frac{h_{H_1,L_1} - h_{H_2,L_2}}{2},
\label{eq:SClim}
\ee
%
%\be
% \< \CO_{H_1}(0) \CO_{H_2}(\infty) \CO_{L_1}(1) \CO_{L_2}(z) \> \ \ \ &\mathrm{with}& \ \ \ \frac{h_{H_i}}{c}, h_{L_i}, \delta_H\textrm{ fixed}, \nn\\
%\delta_{H} \equiv \frac{h_{H_1} - h_{H_2}}{2} &\qquad&  \delta_{L} \equiv \frac{h_{L_1} - h_{L_2}}{2},
%\label{eq:SClim}
%\ee
where $h_{H_i}$ and $h_{L_i}$ are the holomorphic dimensions of $\CO_{H_i}$ and $\CO_{L_i}$.  As suggested in figure \ref{fig:HeavyLightCorrelatorOnAdSCylinder}, via AdS/CFT we would naturally interpret correlators in this limit in terms of the motion of a light probe  interacting with a heavy object or a BTZ black hole \cite{BTZ}.  In the AdS description, the heavy operator produces a classical background in which the light probe moves.

We will argue that such an interpretation also exists in the CFT, and that it can be used as the basis of a precise new computational method.  Specifically, we will show that the computation of Virasoro conformal blocks can be trivialized by placing the CFT in a non-trivial background metric, related to the flat metric by a conformal transformation.

Our main result is that the full Virasoro conformal block for a semi-classical heavy-light correlator is equal to a global conformal block with the light operators evaluated in a new set of coordinates, $\CO_L(z) = (w'(z))^{h_L} \CO_L(w(z))$.  That is, 
\be
\CV(c, h, h_i, z) = \alpha^{-h}(w'(1))^{h_{L_1}} (w'(z))^{h_{L_2}} G(h, h'_i, w), 
\ee
%{\red I think the $\alpha^{-h}$ factor was missing, and is present due to (\ref{eq:newThreePoint})}
where
\be
&&w = z^\alpha  \ \ \ \ \mathrm{with}  \ \ \ \ \alpha = \sqrt{1 -  \frac{24 h_H}{c}},\nn\\
\textrm{ and }&&h_L' = h_L, \quad h_H'=h_H, \quad \delta_L' =\delta_L, \quad \delta_H' = \frac{\delta_H}{\alpha},
\ee
and $G$ is the global conformal block \cite{DO1,DO2,DO3}
\be
G(h, h_i, x) &=& (1-x)^{h - 2h_L}  {}_2F_1(h - 2\delta_H, h + 2\delta_L, 2h, 1-x).
\ee
This result 
confirms and extends the analysis of \cite{Fitzpatrick:2014vua} beyond the range of validity of the older methods we used there.\footnote{In \cite{Fitzpatrick:2014vua} we obtained the Virasoro blocks as $e^{\frac{c}{6} f + g}$, where we computed $f$ but $g$ was left undetermined; in this work we fully determine $g$ up to $1/c$ corrections, which might be computed using recursion relations \cite{ZamolodchikovRecursion,Zamolodchikov:1987}.}   Note that $\alpha$ is related to the Hawking temperature $T_H$ of a BTZ black hole by 
\be
\label{eq:HawkingTemperature}
T_H = \frac{|\alpha|}{2 \pi}
\ee 
when $h_H > \frac{c}{24}$.  

In the next section we will briefly review conformal blocks and the computation of the Virasoro blocks at large central charge.  Then in section \ref{sec:HeavyOperatorsClassicalBackgrounds} we explain our main result, the new recursion relations that use it as a seed, and an extension to include $U(1)$ currents.  In section \ref{sec:AdSInterpretation} we discuss the AdS interpretation of our results, while in section \ref{sec:ImplicationsThermodynamics} we discuss the implications for thermodynamics.  We conclude with a discussion of future directions in section \ref{sec:FutureDirections}.

\section{Review of CFT$_2$ and Virasoro conformal blocks} 
  
In any CFT, correlation functions can be written as a sum over exchanged states
\be
\< \CO_H(\infty) \CO_H(1) \CO_L(z,\bar{z}) \CO_L(0) \> = \< \CO_H(\infty) \CO_H(1) \left( \sum_{\alpha} | \alpha \> \< \alpha | \right)  \CO_L(z,\bar{z}) \CO_L(0) \> .
\ee
We can organize this into a sum over the irreducible representations of the conformal group, which are called conformal blocks.  When working in radial quantization we diagonalize the dilatation operator $D$ and angular momentum generators, so the conformal blocks are labeled by a scaling dimension and by angular momentum quantum numbers.  

In the case of CFT$_2$ the full conformal algebra is two copies of the Virasoro algebra
\be \label{eq:VirasoroAlgebra}
[ L_n, L_m ] = (n-m) L_{n+m} + \frac{c}{12} n (n^2-1) \delta_{n,-m}
\ee
one for the holomorphic ($L_n$) and one for anti-holomorphic ($\bar L_n$) conformal transformations.  The parameter $c$ is the central charge of the CFT. It does not appear in commutation relations of the `global' sub-algebra, 
\be
L_{\pm 1}, L_0, \ \ \ \mathrm{and} \ \ \ \bar L_{\pm 1}, \bar L_0,
\ee
which together form an $SL(2,\mathbb{C})/\mathbb{Z}_2 \cong SO(3,1)$ algebra.  Because of the structure of (\ref{eq:VirasoroAlgebra}), in the limit of $c\rightarrow \infty$ with all other parameters held fixed the Virasoro conformal algebra essentially reduces to the global conformal algebra, as we will review below.  

 The Virasoro conformal blocks are most naturally labeled by holomorphic and anti-holomorphic dimensions, $h$ and $\bar h$, which are the eigenvalues of $L_0$ and $\bar L_0$.  The quantities $h$ and $\bar h$ are restricted to be non-negative in unitary CFTs.  The scaling dimension is $h + \bar h$ and angular momentum is $ h - \bar h $.  Any CFT$_2$ correlation function can be written as a sum over Virasoro conformal blocks, with coefficients determined by the magnitude of matrix elements
\be
\< \CO_H \CO_H | h, \bar h \> \ \ \  \mathrm{and} \ \ \ \< h, \bar h | \CO_L \CO_L \>
\ee
where $| h, \bar h \>$ is a primary state.  Primaries are lowest weight under Virasoro, so $L_n | h , \bar h \> = 0$ for $n > 0$, and similarly for $\bar L_n$.  By the operator state correspondence $\CO_{h, \bar h}(0) | 0 \> = | h , \bar h  \>$, and these matrix elements are just the 3-pt correlation functions of primary operators, which are otherwise known as operator product expansion (OPE) coefficients.

\begin{figure}[t!]
\begin{center}
\includegraphics[width=0.95\textwidth]{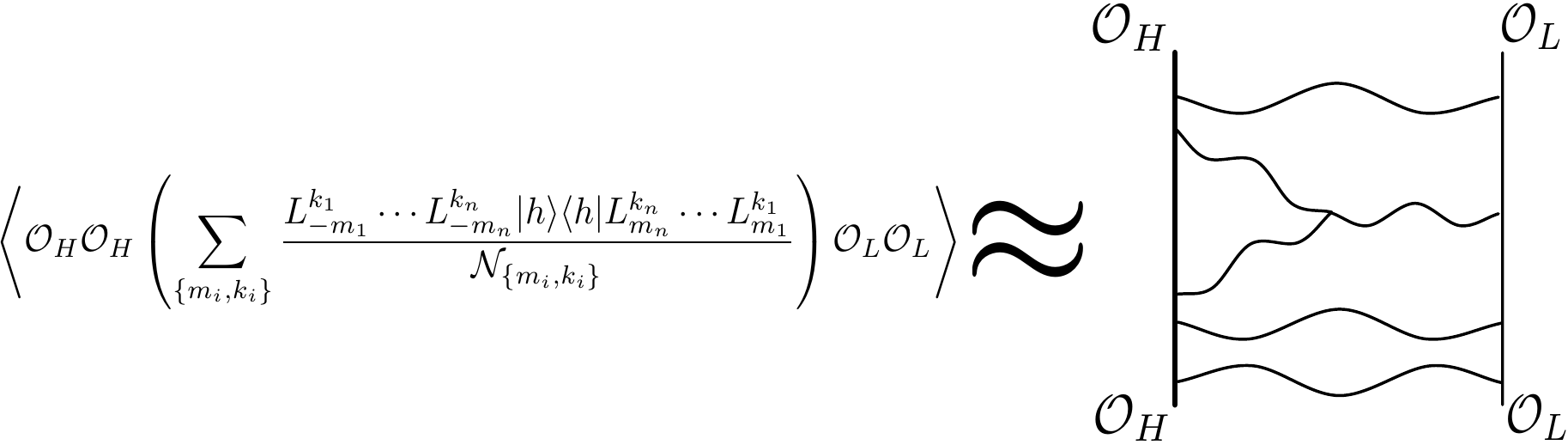}
\caption{ This figure suggests how the exchange of all Virasoro descendants of a state $| h \>$ corresponds to the exchange of $| h \>$ plus any number of gravitons in AdS$_3$.  This is sufficient to build the full, non-perturbative AdS$_3$ gravitational field entirely from the CFT$_2$.
}
 \label{fig:VirasoroDescendantsAsGravitons} 
\end{center}
\end{figure}

The Virasoro generators $L_n$ appear directly in the Laurent expansion of the holomorphic part of the stress energy tensor 
\be
T(z) = \sum_n z^{-n-2} L_n,
\ee
 with $T(z) = -2 \pi T_{zz}(z)$ and similarly for the anti-holomorphic $\bar T(\bar{z}), \bar{L}_n$, and  $T_{\bar{z} \bar{z}}$.  The Virasoro algebra can be derived directly from the singular terms in the OPE of $T(z) T(0)$.  To take the adjoint of an operator we perform an inversion $z \to 1/\bar{z}$, because we are working in radial quantization.  This means  
\be
[T(z)]^\dag = \frac{1}{\bar{z}^4} T(1/\bar{z}),
\ee
which leads to the relation $L_n^\dag = L_{-n}$ for the Virasoro generators.  Here we have assumed that the CFT$_2$ lives in the plane, with metric $ds^2 = dz d \bar z$, as is conventional.

\subsection{Relation to AdS/CFT}

Via the AdS/CFT correspondence, the CFT stress tensor (and the $L_n$ with $n \leq -2$) creates gravitons in AdS$_3$.  The central charge determines the strength of gravitational interactions via the relation $c = \frac{3}{2 G}$, where $G$ is the AdS Newton constant.  In global coordinates, the AdS Hamiltonian is the dilatation operator
\be
D = L_0 + \bar L_0,
\ee
and so CFT scaling dimensions determine AdS energies, with the Planck scale corresponding to a dimension $\sim \frac{c}{12}$ in the CFT.  

Correlation functions in the CFT can be given an AdS interpretation, as indicated in figure \ref{fig:HeavyLightCorrelatorOnAdSCylinder}.  The Virasoro conformal blocks encapsulate all gravitational effects in AdS$_3$, as suggested by figure \ref{fig:VirasoroDescendantsAsGravitons}.  As we reviewed extensively in \cite{Fitzpatrick:2014vua}, CFT states with dimension $h_H \propto c$ correspond to AdS$_3$ objects that produce either deficit angles or BTZ black holes.  We will study the Virasoro blocks in the limit
\be
 h_H, \bar h_H  \propto c \ \gg \  h_L , \bar h_L,
\ee
which we will refer to as the heavy-light semi-classical limit.  In this limit AdS gravity is weakly coupled, but the heavy operator $\CO_H$ has a Planckian mass.  The operator $\CO_L$ creates a state that can be thought of as probing the heavy AdS deficit angle or black hole. We emphasize that our computations will all be completely independent of AdS, but AdS/CFT provides the most natural interpretation of and motivation for our results.

\subsection{Virasoro blocks at large central charge}
\label{sec:BasicVirasoroBlocksLargeC}

Here we review the well-known fact that the Virasoro blocks simplify dramatically if we take the limit $c \to \infty$ with other operator dimensions fixed, reducing to representations of the much smaller global conformal group.  Specifically, consider the correlator
\be
\< \CO_L(\infty)  \CO_L(0) \CO_L(1) \CO_L(z, \bar z) \>
\ee
for the primary operator $\CO_L$ with dimensions $h_L, \bar h_L$ in the limit that
\be
c \to \infty  \ \ \ \mathrm{with} \ \ \  h_L, \bar h_L \ \ \mathrm{fixed}.
\ee
In this case the holomorophic Virasoro blocks are simply
\be \label{eq:GlobalBlock}
G_{h}(z) = (1-z)^{h-2h_L} {}_2 F_1 \left(h, h, 2h, 1-z \right) + O \left( \frac{1}{c} \right),
\ee
 where we assume that $h$, the intermediate holomorphic dimension, is also fixed as $c \to \infty$.  
That is, the Virasoro conformal block reduces to a conformal block for the global conformal algebra, generated by $L_{\pm 1}$ and $L_0$ alone, in the large $c$ limit.  

To see why this is the case, consider the projection operator $\CP_h$ onto the states in Virasoro conformal block.  The basis of states
\be
L_{-m_1}^{k_1} \cdots L_{-m_n}^{k_n} | h \>
\ee
is approximately orthogonal to leading order in $1/c$.  Thus we can approximate the projection operator onto the conformal block as
\be
\label{eq:VirasoroProjector}
\CP_{h} \approx \sum_{\{m_i,k_i\}} \fr{L_{-m_1}^{k_1} \cdots L_{-m_n}^{k_n} | h \>\< h | L_{m_n}^{k_n} \cdots L_{m_1}^{k_1}}{\CN_{\{m_i,k_i\}}}.
\ee
In this expression we are summing over all possible states created by acting on the primary state $| h \>$ with any number of Virasoro generators, and dividing by their normalization.  Using the Virasoro algebra from equation (\ref{eq:VirasoroAlgebra}) it is easy to see that the normalization
\be
\< h | L_{n} L_{-n} | h \> = \left( 2 n h + \frac{c}{12} n(n-1)(n+1) \right)
\ee
for $n > 0$, and so for $n \geq 2$ the normalization is of order $c$.  The $L_{-m}$ in the numerator of $\CP_h$ can produce powers of $h$ and $h_L$ via the relation
\be \label{eq:VirasoroPrimaryAction}
\comm{L_{m}}{\CO(z)} = h_i(1+m) z^{m} \CO(z) + z^{1+m} \partial_z \CO(z)
\ee
for primary operators $\CO$, but the numerator of $\CP_h$ will be independent of $c$.  This means that the normalizations $\CN_{\{m_i,k_i\}}$ are proportional to a polynomial in $c$ if any $m_i \geq 2$, and that there are no factors in the numerator to make up for this suppression, so terms with $m_i \geq 2$ make negligible contributions to the conformal block at large $c$.  The projector reduces to
\be
\label{eq:GlobalProjector}
\CP_{h} \approx  \sum_{k} \fr{ L_{-1}^{k} | h \>\< h | L_{1}^{k}}{\< h | L_1^k L_{-1}^k | h \>}  + O \left( \frac{1}{c} \right),
\ee
which just produces the 2d global conformal block for the holomorphic sector.  The result is equation (\ref{eq:GlobalBlock}).  Readers interested in a detailed review of all these statements can consult appendix B of \cite{Fitzpatrick:2014vua}.

\begin{figure}[t!]
\begin{center}
\includegraphics[width=0.95\textwidth]{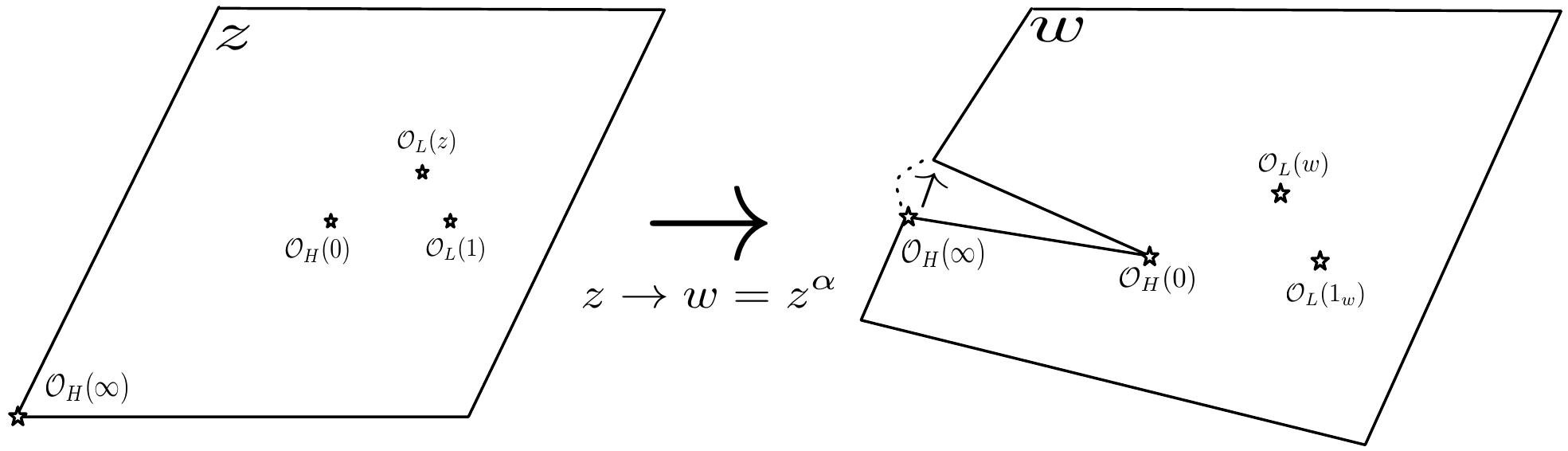}
\caption{ This figure suggests the conformal mapping from $z$ to $w$ coordinates, so that stress-tensor exchange between heavy and light operators has been absorbed by the new background metric. Note that we have placed the heavy operators at $0$ and $\infty$ for simplicity, although we place one of them at $1$ to emphasize a particular OPE limit in section \ref{sec:HeavyOperatorsClassicalBackgrounds}. }
 \label{fig:ConformalMappingZtoW} 
\end{center}
\end{figure}

\section{Heavy operators and classical backgrounds for the CFT}
\label{sec:HeavyOperatorsClassicalBackgrounds}

\subsection{Stress tensor correlators and the background metric}
\label{sec:BackgroundMetric}

The large central charge analysis above becomes invalid if the dimensions of the various operators are not fixed as $c \to \infty$. When the dimensions become large, the numerator of $\CP_h$ in equation (\ref{eq:VirasoroProjector}) can compensate for the powers of $c$ in the denominator.  The problem is that one obtains contributions
\be
\< \CO_H \CO_H L_{-m_1}^{k_1} \cdots L_{-m_n}^{k_n} | h \>
\ee
that are proportional to some polynomial in $h_H$, due to the relation of equation (\ref{eq:VirasoroPrimaryAction}), and $h_H \propto c$.  Since the $L_n$ are just Laurent coefficients in the expansion of the stress tensor, we can reformulate the problem in terms of the correlators
\be
\< \CO_H \CO_H T(x_1) \cdots T(x_k) \CO_h(0) \>.
\ee
The issue is that this correlator will involve powers of the dimension $h_H$, and we would like to take $h_H \propto c$ as $c \to \infty$.

The crucial idea behind our method is that \emph{we can eliminate all dependence on $h_H$ by performing a conformal transformation to a new background.}  A well-known feature of the CFT$_2$ is that the stress tensor does not transform as a primary operator.  Physically, this can be regarded as a consequence of the fact that $T$ picks up a vacuum expectation value (VEV) when the CFT is placed in a non-trivial 2d background metric $g_{\mu \nu}$.  We will choose $g_{\mu \nu}$ so that the VEV of $T$ cancels its large matrix element with $\CO_H$.  The relevant mapping is illustrated in figure \ref{fig:ConformalMappingZtoW}.

Recall that in a CFT, conformal transformations act on primary operators as
\be \label{eq:PrimaryTransformation}
\CO(z) \to \CO(w) \equiv \left[ \frac{dz}{dw} \right]^{h_\CO} \CO(z(w))
\ee 
under a conformal transformation $z \to w(z)$.
The conformal transformation rule for the stress tensor $T(z)$ is
\be \label{eq:TTransformationRule}
T(z) \to T(w) = \left[ \frac{dz}{dw} \right]^2 T(z(w)) + \frac{c}{12} S(z(w), w).
\ee
The first term is the transformation rule for a primary operator, while the second term 
\be \label{eq:SchwarzianDefinition}
S(z(w),w) \equiv \frac{z'''(w) z'(w) - \frac{3}{2} (z''(w))^2}{(z'(w))^2}
\ee
is the Schwarzian derivative. 

A physical way to think about the transformation rule (\ref{eq:TTransformationRule}) is that $T$ obtains a VEV in a background metric.  We can use the result that in a general background
\be
ds^2 = e^{2 \sigma(z, \bar z)} dz d \bar z
\ee
the stress tensor picks up the vacuum expectation value \cite{Brown:1977sj, Herzog:2013ed}   
\be
\< T^{\mu \nu} \> = \frac{c}{12 \pi} \left( \nabla^{\mu} \nabla^{\nu} \sigma + \nabla^\mu \sigma \nabla^\nu \sigma - g^{\mu \nu} \left( \nabla_\lambda \nabla^\lambda \sigma + {\frac{1}{2}}\nabla_\lambda \sigma \nabla^\lambda \sigma \right) \right).
\ee
%{\red Missing factor of 1/2?}
After the conformal transformation $z \to w(z)$  the CFT lives in a background metric 
\be
ds^2 = w'(z) dz \bar dz.
\ee
By using $w'(z) = e^{2 \sigma}$  and the normalization $T(z) = 2 \pi T_{zz}(z)$, we obtain the Schwarzian derivative in equation (\ref{eq:SchwarzianDefinition}) from the general formula for the VEV. 

Now let us demonstrate how we can make use of these results.  We would like to study the correlator
\be
\< \CO_H(\infty) \CO_H(1) T(z) \CO_h(0) \> = C_{HHh} \left( \frac{h_H}{(1-z)^2} + \frac{h}{(1-z) z^2} \right),
\ee
where $C_{HHh}$ is an OPE coefficient. Note that we are now placing the second heavy operator at $z=1$ rather than $z=0$ as before.  This will cause less clutter in many expressions since the OPE limit will correspond to an expansion in powers of $z$ rather than powers of $1-z$.  If we perform a conformal transformation to $w(z)$ defined by
\be \label{eq:wDefinition}
1-w = (1-z)^\alpha \ \ \ \mathrm{with}  \ \ \ \alpha = \sqrt{1 - 24 \frac{h_H}{c} },
\ee
then in the new coordinate system 
\be \label{eq:TransformedHeavyCorrelator}
\< \CO_H(\infty) \CO_H(1) T(w) \CO_h(0) \> = C_{HHh} \left( h \,  \frac{ 1-z(w) }{z^2(w)} \right).
\ee
The key point is that all $h_H$ dependence has been removed through a cancellation between the Schwarzian derivative and the primary transformation law.  The conformal transformation itself could have been determined a priori by writing the result for a transformation parameterized by a general $w(z)$ and then solving a differential equation demanding that all $h_H$ dependence be eliminated.  In fact, one can generalize the procedure to find more complicated conformal transformations that cancel the contribution from more than two heavy operators.

We have seen how operators transform, but now we would like to study the states that they create, especially the Virasoro descendants or `gravitons' created by the stress tensor.  We can expand $T(w)$ in the new coordinate $w$ 
\be
T(w) = \sum_n w^{-2-n} \CL_n.
\ee
The Virasoro algebra can be derived entirely from the singular terms in the $T(z)T(0)$ OPE, and these terms are preserved by conformal transformations.  An important consequence is that the new generators $\CL_n$ expanded in $w$ still satisfy the usual Virasoro algebra from equation (\ref{eq:VirasoroAlgebra}), with $L_n \to \CL_n$, and also the relation (\ref{eq:VirasoroPrimaryAction}) when $\CL_n$ act on conformally transformed operators $\CO(w)$.  

The $\CL_n$ are a complete basis of Virasoro generators, so one can write $L_m$ as a linear combination of the $\CL_n$ and vice versa.   A non-trivial special feature of the $\CL_n$ generators is
\be
\CL_n | h \> = 0 \ \ \ \mathrm{for}  \ \ \ n \geq 1
\label{eq:LprimeLowest}
\ee
when the state $| h \>$ is primary. Consequently, all states in the representation are generated by acting with $\CL_{-n}$'s, $n>0$.  
 These statements depend on the fact that the inverse transformation
 $z(w)$ has an analytic Taylor series expansion in $w$ when expanded about the origin, as can be seen from equation (\ref{eq:wDefinition}). Therefore, because correlators of the form $\< \alpha | T(z) |h\>$ are regular as $z\rightarrow 0$ for all $\< \alpha|$, they are also regular in $w$ coordinates  $\< \alpha | T(w) |h\>$ as $w\rightarrow 0 $.  Equation (\ref{eq:LprimeLowest}) follows directly.

We can also be more explicit and use the transformation rule from equation (\ref{eq:TTransformationRule}) to write the $w$-expansion of $T(w)$ directly in terms of $T(z(w))$, which gives the relation
\be
\sum_n w^{-n-2} \CL_{n} = \frac{1}{\alpha} \left( 1-w \right)^{\frac{1}{\alpha}-1} \sum_m [z(w)]^{-m-2} L_{m} - \frac{h_H}{\alpha^2(1-w)^2}.
\ee
All terms on the right-hand side have a series expansion in positive integer powers of $w$, so when we combine terms proportional to $w^{-n-2}$, we will include only $L_{m}$ with $m \geq n$.  This provides an alternate and more explicit derivation of equation (\ref{eq:LprimeLowest}).

The conformal transformation $z(w)$ has a branch cut running to infinity, so it cannot be expanded in a Taylor series around $w = z = \infty$.  As a result the $\CL_n$ do not have well-behaved adjoints with respect to inversions.  In particular, unlike the case of the usual Virasoro generators
\be 
\CL_n^\dag \neq \CL_{-n}
\ee
so the $\CL_n$ do not act to the left on `ket' states $\< h |$ in a simple way.  Nevertheless, 
we may formally define the `$w$-vacuum state' $\< 0_w |$, chosen to satisfy
\be
\< 0_w | \CL_{-n} = 0
\ee
for all $n \geq -1$, and we can normalize it so that $\< 0_w | 0 \> = 1$.  Similarly, for any primary state $| h \>$ there exists a ket $\< h_w |$ defined through
\be \label{eq:Definitionhw}
\< h_w |     = \lim_{w \to \infty} \< 0_w | w^{2h} \CO(w).
\ee
Since the new Virasoro generators $\CL_n$ act on operators in the new metric such as $\CO(w)$ in exactly the same way that the conventional generators act on operators in a flat metric, matrix elements can be easily calculated.  
The $\< h_w | $ state  satisfies
\be
\< h_w | \CL_0 = \< h_w | h \ \ \  \mathrm{and}  \ \ \ \< h_w | h \>  = 1
\ee
so for most purposes we can use $\< h_w |$ in place of $\< h |$, and $\CL_n$ in place of $L_n$. We will use this construction in the next section to create a modified version of the Virasoro conformal block projector from equation (\ref{eq:VirasoroProjector}), utilizing the $\CL_n$ in place of the $L_n$ generators.

\subsection{Heavy-light Virasoro blocks from a classical background}

We would like to compute the conformal blocks for the semi-classical heavy-light correlator 
\be
\CV_h(z) = \< \CO_{H_1}(\infty) \CO_{H_2} (1)  \left[ \CP_h  
\right]   \CO_{L_1} (z) \CO_{L_2} (0) \> 
\ee
where $\CP_h$ projects onto the state $| h \>$ and all of its Virasoro descendants.  We will choose the normalization of $\CV_h(z)$ so that its leading term is $z^{h- 2h_L}$ in an expansion around $z=0$.  We now allow $h_{H_1} \neq h_{H_2}$ and $h_{L_1} \neq h_{L_2}$ for maximal generality, but we will need to take 
\be
\delta_H = h_{H_1} - h_{H_2} \ \ \ \mathrm{with} \ \ \ | \delta_H | \ll c
\ee
and we also define $\delta_L = h_{L_1} - h_{L_2}$.  This computation is naively much more difficult than the one in section \ref{sec:BasicVirasoroBlocksLargeC}, because the Virasoro generators act on $\CO_H$ to produce factors of $h_H \propto c$.  This means that the contribution from states containing $L_{-n}$ with $n>1$ in $\CP_h$ no longer vanish, since the factors of $h_H$ in $\< \CO_{H_1}(\infty) \CO_{H_2}(1) L_{-n} | h\> $ compensate for factors of $c$ in the normalization.  However, we will make use of the conformal transformation from the previous section to simplify the computation, basically turning it into a recapitulation of the global conformal case.

Let us begin by transforming the light operators in the correlator to $w$ coordinates, so we want to compute the Virasoro blocks for
\be
\CV_h(w) = \< \CO_{H_1}(\infty) \CO_{H_2} (1)  \left[ \CP_h 
\right]   \CO_{L_1} (w) \CO_{L_2} (0) \>.
\ee
This formula differs from $\CV_h(z)$ by the Jacobian factor $(z'(w))^{h_{L_1}} (z'(0))^{h_{L_2}}$, which we will restore below.  We claim that in the heavy-light semi-classical limit
\be
\label{eq:NewVirasoroProjector}
\CP_{h,w} \approx \sum_{\{m_i,k_i\}} \fr{\CL_{-m_1}^{k_1} \cdots \CL_{-m_n}^{k_n} | h \>\< h_w | \CL_{m_n}^{k_n} \cdots \CL_{m_1}^{k_1}}{\CN_{\{m_i,k_i\}}} 
\ee
is a projector onto the irreducible representation of the Virasoro algebra with primary $| h \>$.  We have used $\approx$ because this basis is not orthogonal at higher orders in $1/c$, though one can easily write down a formula without approximations by using the matrix of inner products of states instead of just using the normalizations $\CN_{\{m_i, k_i\}}$.  We need to show that the normalizations for $\CL_n$ are identical to those for $L_n$, that this is a projection operator satisfying $(\CP_{h,w})^2 = \CP_{h,w}$, and that it is complete, i.e. that all states in the representation are included with proper normalization.  The state $\< h_w |$ was defined in equation (\ref{eq:Definitionhw}).   

If $\CP_{h, w}$ is a projector, then its completeness follows because the $\CL_n$ form a basis for the Virasoro algebra.  The fact that $\CP_{h,w}$ is projector, and the equivalence of normalization factors, follows from the general identification of matrix elements
\be
\< h_w | \CL_{m_n}^{k_n} \cdots \CL_{m_1}^{k_1} | h \> = \< h | L_{m_n}^{k_n} \cdots L_{m_1}^{k_1} | h \>  
\label{eq:LwMatrixElements}
\ee
for arbitrary $m_i$, $k_i$, and $n$.  This identification holds because the $\CL_m$ and $L_m$ both satisfy the Virasoro algebra, so we can use their commutation relations to move $m_i < 0$ to the left and $m_i > 0$ to the right.  Since both generators annihilate $| h \>$ when $m_i > 0$ and $\< h_w|$ and $\< h |$ when $m_i < 0$, the full computation is determined by the Virasoro algebra.  Note that despite the presence of the adjoint state $\< h_w |$ we have not made use of $\CL_n^\dag$.\footnote{One may worry whether the state $\<h_w|$ is physical and thus whether our construction is meaningful.  Ultimately, one may view it simply as a convenience for constructing the projection operator $\CP_{h,w}$, whose action on any state can be defined operationally by using (\ref{eq:LwMatrixElements}).  The projector thus defined is easily seen to satisfy $\CP_{h,w}^2 = \CP_{h,w}$, and also to project any state onto the basis $\CL_{-m_1}^{k_1} \dots \CL_{-m_n}^{k_n}|h\>$, which is all that is required for the construction of the conformal blocks.} 

Thus we can compute the heavy-light Virasoro block using
\be
\CV_h(w) = \< \CO_{H_1}(\infty) \CO_{H_2} (1)  \left[ \CP_{h,w} 
\right]   \CO_{L_1} (w) \CO_{L_2} (0) \>. 
\ee
The matrix elements 
\be
\< h_w | \CL_{m_n}^{k_n} \cdots \CL_{m_1}^{k_1} \CO_{L_1} (0) \CO_{L_2} (w) \> 
\ee
are identical to the matrix elements of light operators in a flat background, except now they depend on $w$ instead of $z$.  This again follows because the $\CL_n$ satisfy the Virasoro algebra, and act according to equation (\ref{eq:VirasoroPrimaryAction}) except with $L_n \to \CL_n$ and $\CO(z) \to \CO(w)$.  However, the heavy matrix elements 
\be
\< \CO_{H_1}(\infty) \CO_{H_2} (1) \CL_{-m_1}^{k_1} \cdots \CL_{-m_n}^{k_n} | h \>
\ee
are non-trivial, and have no direct equivalent in a flat background.  They can be obtained from equation (\ref{eq:TransformedHeavyCorrelator}), and its generalization to multiple $T(w)$ insertions, simply by expanding in $w$.  In fact, in the case of the global generator $\CL_{-1} = \partial_w$, we need only compute
\be
\<\CO_{H_1}(\infty) \CO_{H_2}(1) \CL_{-1}^k |h\> &=& \lim_{w\rightarrow 0} \partial_w^k \<\CO_{H_1}(\infty) \CO_{H_2}(1) \CO_h(w) \>  \nn\\
 &=&\alpha^{-h} c_{H_1 H_2 h} \lim_{w \rightarrow 0} \partial_w^k (1-w)^{-h +  \delta_H/\alpha}
 \label{eq:newThreePoint}
 \ee
since $\CL_{-1}$ translates $\CO_h$ in the $w$ coordinate system.  This differs from the equivalent result in a flat metric only by the intriguing replacement $\delta_H \to \delta_H/\alpha$.

We have all the ingredients we need to compute the heavy-light Virasoro block at large central charge.  Just as in section \ref{sec:BasicVirasoroBlocksLargeC}, because the normalizations $\CN_{\{m_i,k_i\}}$ are proportional to positive powers of $c$ when $m_i \neq \pm 1$, the Virasoro block reduces to
\be
\label{eq:GlobalPrimedProjector}
\CV_{h}(w) =  \<\CO_{H_1}(\infty) \CO_{H_2}(1) \left( \sum_{k} \fr{ \CL_{-1}^{k} | h \>\< h_w | \CL_{1}^{k}}{\< h_w | \CL_1^k \CL_{-1}^k | h \>} \right) \CO_{L_1} (w) \CO_{L_2} (0) \>. 
\ee
Only the heavy side differs from the usual global conformal block, due to the $\delta_H$ rescaling.  Performing the sum, we find that in the scaling limit of (\ref{eq:SClim}),
\be
\lim_{c \rightarrow \infty} \CV(c, h_p, h_i, z) =  (1-w)^{(h_L+\delta_L)(1-\frac{1}{\alpha})} \left( \frac{w}{\alpha} \right)^{h-2h_L} {}_2F_1\left(h - \frac{\delta_H}{\alpha}, h + \delta_L, 2 h, w \right) ,
\label{eq:SCblockFinal}
\ee
%{\red another missing $\alpha^{-h}$? actually it looks like it should be $\alpha^{-h+2h_L}$?}
where we view $w$ as a function of $z$ through equation (\ref{eq:wDefinition}), and we have included the Jacobian factors $w'(z)^{h_{L_i}}$ from relating $\CO_{L_i}(w)$ to $\CO_{L_i}(z)$.  The Virasoro block is normalized so that the leading term at small $z$ is $z^{h-2h_L}$. 
This formula agrees with our less powerful results from \cite{Fitzpatrick:2014vua}, and we have checked up to the first six terms that it matches a series expansion of the exact Virasoro block in this limit, computed using recursion relations derived by Zamolodchikov \cite{ZamolodchikovRecursion,Zamolodchikov:1987}.

\subsection{Order of limits at large central charge}
\label{sec:OrderofLimits}

 The key advance beyond \cite{Fitzpatrick:2014vua} is that previously, only part of the conformal block was known in the heavy-light large $c$ limit, whereas (\ref{eq:SCblockFinal}) contains the full answer at $c \rightarrow \infty$.   The results in \cite{Fitzpatrick:2014vua} apply to a different regime of parameter space, so let us clarify how those results relate to the present ones.

The `monodromy method' we employed in \cite{Fitzpatrick:2014vua} formally only applies to the limit where all $h_i/c$ and $h_p/c$ are held fixed while $c\rightarrow \infty$. In this limit, the conformal blocks are expected to grow like $\CV \sim e^{c f(z) } g(z)$ for some function $f$ that depends only on the ratios $h_i/c, h_p/c$.  Such methods cannot determine the $c$-independent prefactor $g(z)$.   Furthermore, it is very challenging to determine the function $f(z)$ exactly for arbitrary $h_i/c, h_p/c$, and in practice additional approximations must be made (but see \cite{Painleve6} for further ideas).  Consequently, we made the further expansion of $f$ in small $h_L/c$ and worked to linear order only, while working exactly as a function of $h_H/c$.  

In fact, it is not immediately obvious that the limit (\ref{eq:SClim}) considered in this paper has any overlap with that just described, i.e. that the limits 
\be
 \lim_{h_L/c \rightarrow 0} \frac{c}{h_L}\left[ \lim_{c\rightarrow \infty, h_L/c \textrm{ fixed }} \frac{1}{c} \log \frac{\CV(z)}{z^{h_p-2h_L}}\right] = 
\lim_{h_L \rightarrow \infty} \frac{1}{h_L} \left[ \lim_{c\rightarrow \infty, h_L \textrm{ fixed }} \log \frac{\CV(z)}{z^{h_p-2h_L}} \right]
\ee
(with $h_H/c$ and $h_p/h_L$ fixed on both sides) are equivalent.  In particular, one might wonder why coefficients like $\frac{h_H h_p + h_p^3}{c  + h_p^2}$, which would give $\frac{h_H}{c} \frac{h_p}{h_L}$ in the second limit but $\frac{h_p}{h_L}$ in the first limit, never appear.

To understand why the  limits are equivalent, one can compare terms in $\log \CV$ in a series expansion in $z$, each of whose coefficients is a rational function of $h_L$ and $c$.
At each level, the Virasoro block projectors produce rational functions of $c,h_p$ where the denominator comes solely from inverting the Gram matrix. This means that the denominator is proportional to the Kac determinant, $\prod_{m, n} (h_p-h_{m,n}(c))$, where $h_{m,n}(c) = \frac{c}{24}((m+n)^2-1)+ \CO(c^0)$ and the product is taken over all $m,n$ corresponding to the Gram matrix at that level.  

At large $c$ in either limit ($h_p$ fixed or $h_p/c$ fixed), the leading terms in the denominator all scale like the same power of $s$ under $c\rightarrow c s, h_p \rightarrow h_p s$, so coefficients like $\frac{h_H}{c} \frac{h_p}{h_L}$ are indeed impossible. In more physical terms, in this limit we can think of $c$ and $h_p$ having mass dimension, so that the form of the result is constrained by dimensional analysis. 

Consequently, the denominator is a polynomial in $h_p$ and $c$, and the leading term at large $c$ with $h_p/c$ fixed is $\prod_{m, n} \left(h_p - \frac{c}{24}((m+n)^2-1)\right)$, and the $h_p$ is formally negligible if we fix $h_p$ in the large $c$ limit.  Taking the logarithm of the Virasoro block mixes powers of $z$ at many different orders, but 
 the denominator of each resulting term is a product of the original denominators  and thus does not change the fact that both limits have the same leading term.  The leading term in the numerator must be the same because both the $h_p/c$ fixed and $h_p$ fixed limits exist.  The power of $c$ and $h_p$ of the leading term in the denominator fixes the power of $c$ and $h_p$ of the leading term in the numerator.

\subsection{Recursion formula}
\label{sec:RecursionFormula}

Our result (\ref{eq:SCblockFinal}) for the heavy-light conformal block at $c=\infty$ can be systematically improved by adapting the recursion relations of \cite{ZamolodchikovRecursion,Zamolodchikov:1987}.  The basic idea of this method is to consider the conformal blocks as analytic functions of $c$.  Poles in $c$ arise at discrete values $c_{m,n}$ where the representation becomes null, parameterized by pairs of integers $m,n$ satisfying $m\ge 1, n\ge 2$.  The exact conformal block is a sum over its poles plus the value at $c=\infty$:
\be
\CV(c, h_i, h_p, z) = \CV(\infty, h_i, h_p, z) + \sum_{m,n} \frac{R_{m,n}(c_{m,n}(h_p),h_i)}{c-c_{m,n}(h_p)} \CV(c_{m,n}(h_p), h_i, h_p+mn,z) . \quad 
\label{eq:recursion} 
\ee
The residues of the poles must be conformal blocks with arguments shifted as indicated above, leading to a recursion relation seeded by the $c=\infty$ piece.  One can treat the conformal block as an analytic function of $c$ with $h_i,h_p$ fixed, as in the original papers \cite{ZamolodchikovRecursion,Zamolodchikov:1987}, or one can treat it as an analytic function of $c$ with $h_H/c, h_L, \delta_L, \delta_H$ fixed.  In the former case, the $c=\infty$ piece is just the global conformal block, but in the latter case it is the semi-classical block (\ref{eq:SCblockFinal}).  When applying the above formula in this case, it is important to remember that $h_H$ has been promoted to a function of $c$, i.e. $h_H(c)\equiv  \eta_H c$ with $\eta_H$ fixed, so on the RHS we have $h_H \rightarrow h_H(c_{m,n}) = \eta_H c_{m,n}$.  Explicit expressions for the coefficients $R_{m,n}$ were derived in \cite{ZamolodchikovRecursion,Zamolodchikov:1987} and are given in appendix \ref{app:recursion}, along with expressions for $c_{m,n}$.  

Starting from the $c=\infty$ seed in the recursion formula, each subsequent term in the sum involves conformal blocks with the weight of the exchanged primary  operator increased by $mn$.  Since each block is a sum over powers of $w$ beginning with $w^{h_p}$ and increasing by integer powers, this means that only a finite number of terms in (\ref{eq:recursion}) can ever contribute to a given power of $w$.    The recursion formula provides an efficient method for computing the exact series coefficients of the Virasoro blocks in an expansion in powers of $w$.  
  
It is less clear what can be learned analytically in powers of $1/c$.  At leading order in $1/c$, one may neglect the $c_{m,n}$ in the denominator and write  
\be
\CV(c, h_i, h_p, z) = \CV(\infty, h_i, h_p, z) + \frac{1}{c}\sum_{m,n} R_{m,n}(c_{m,n}(h_p),h_i)\CV(c_{m,n}(h_p), h_i, h_p+mn,z)  + \CO(1/c^2). \nn\\
\label{eq:recursion2}
\ee
This expansion should contain universal information about quantum $1/{m_{\rm pl}}$ corrections in gravitational theories.  To obtain non-perturbative information one would need to solve the recursion relation exactly for some $w$ and study the behavior in $1/c$.

\subsection{Extension to $U(1)$} 

In the case where the theory contains additional conserved currents, the symmetry algebra of the theory is enlarged beyond the Virasoro algebra.  The simplest such extension is to add a $U(1)$ current $J$, enhancing the algebra to include a Kac-Moody algebra:
\be
\left[ L_n, L_m \right]  & = & (n-m) L_{n+m} + \frac{c}{12} (n^3 - n) \delta_{ n + m, 0 },
\nn \\
\left[ L_{n} , J_{m} \right]  & = &  - m J_{n+m},
\nn \\
\left[ J_n, J_m \right] &=& n k \delta_{n+m, 0},
\nn\\
\left[ J_n, \phi(z) \right] &=& q_\phi z^n \phi(z),
\ee
where $J(z) = \sum_n z^{-n-1} J_n$ and $k$ is the level of the current.  Now, we want to know the $U(1)$-extended Virasoro conformal blocks $\CV_{T+J}$, i.e. the contribution from irreps of the extended algebra.  To simplify its computation, we choose a new basis of operators as follows.  We start with the well-known construction of the Sugawara stress tensor $T_{\rm sug}$ (see e.g. \cite{CFTbook}) 
\be
T^{\rm sug}(z) &=& \sum_n z^{-n-2} L_n^{\rm sug}, \nn\\
L_n^{\rm sug} &=& \frac{1}{2k}\left( \sum_{m\le - 1} J_m J_{n-m} + \sum_{m\ge 0} J_{n-m} J_m \right)  .
\ee
The Sugawara generators act almost like Virasoro generators with central charge $c=1$:
\be
 \left[L^{\rm sug}_n, L^{\rm sug}_m\right] &=& (n-m) L^{\rm sug}_{n+m} + \frac{1}{12} n(n^2-1) \delta_{n+m,0}, \nn\\
 \left[L_n^{\rm sug}, J_m\right] &=& - m J_{n+m},
 \ee
and a short computation shows 
\be
[L_n, L_m^{\rm sug} ] &=& (n-m) L_{n+m}^{\rm sug} + \frac{1}{12} n(n^2-1)\delta_{n+m,0}.
\ee
 One may think of the true stress tensor as getting a contribution from the $U(1)$ charge through the Sugawara stress tensor, so that by subtracting out this contribution we can factor out the contributions to the extended conformal blocks from the current.  More precisely, we define
 \be
 L^{(0)}_n &\equiv& L_n - L^{\rm sug}_n.
 \ee
The key point is that the $L^{(0)}_n$ and $J_n$ generators provide a basis that factors the algebra into separate Virasoro and $U(1)$ sectors:
\be
\left[ L^{(0)}_n, L^{(0)}_m \right]  & = & (n-m) L^{(0)}_{n+m} + \frac{c-1}{12} n(n^2 - 1) \delta_{ n + m, 0 },
\nn \\
\left[ L^{(0)}_{n} , J_{m} \right]  & = &  0. 
\ee 
Furthermore, it is straightforward to check that states $|\phi\>$ that are primary with respect to the $L_n$'s and $J_n$'s, with weight $h$ and charge $q_\phi$, are primary under $L_n^{(0)}$ as well, with weight 
\be
h^{(0)}= h - \frac{q_\phi^2}{2k}.
\ee
In the conformal block, all descendants of the exchanged state $|h_p\> $ can be written in a basis of $J_n$'s and $L_n^{(0)}$s.  We restrict our attention to conformal blocks where the exchanged primary is neutral, i.e. $q_{H_1}= - q_{H_2} \equiv q_H, q_{L_1} = -q_{L_2} \equiv q_S$.    Consider the action of any of the $L_n^{(0)}$s inside the three-point function $\< h_p| \phi_{L_1}(z) \phi_{L_2}(0)\>$:
\be
\< h_p | L_n^{(0)} \phi_{L_1}(z) \phi_{L_2}(0) \> &=& \< h_p | \left(L_n - \frac{1}{2k} \left(J_1 J_{n-1} + J_2 J_{n-2} + \dots J_{n-1}J_1 \right) \right) \phi_{L_1}(z) \phi_{L_2}(0)\>,\nn\\
\ee
where we have used the fact that $J_n$ with $n\le 0$ annihilates $\< h_p|$.  
Now, note that
\be
\< h_p| \phi_{L_1} (z) \phi_{L_2}(0)\> &=& c_{L_1 L_2 p} \frac{1}{z^{h_{L_1}+ h_{L_2} - h_p}} = c_{L_1 L_2 h_p} \frac{z^{-q_L^2/k}}{z^{h_{L_1}^{(0)} + h_{L_2}^{(0)} - h_p}}.
\ee
Thus, we find that $L_n^{(0)}$ acts to the right like $L_n$, but with $h_{L_1}, h_{L_2}$ replaced with $h_{L_1}^{(0)}, h_{L_2}^{(0)}$:
\be
\< h_p | L_n^{(0)} \phi_{L_1}(z) \phi_{L_2}(0) \> &=& \< h_p | (h_{L_1}(n+1) z^n + z^{n+1} \partial_z - \frac{n-1}{2k} q_L^2 z^n) \phi_{L_1}(z) \phi_{L_2}(0) \> \nn\\
&=& z^{-q_L^2/k} (h_{L_1}^{(0)}(n+1) z^n + z^{n+1} \partial_z ) c_{L_1 L_2 h_p} \frac{1}{z^{h_{L_1}^{(0)} + h_{L_2}^{(0)} - h_p}} \nn\\
 &=&z^{-q_L^2/k} \left[ \< h_p | L_n \phi_{L_1} (z) \phi_{L_2}(0)\> \right]_{h_{L_i} \rightarrow h_{L_i}^{(0)}}.
\ee
Descendants with multiple $L_n^{(0)}$s follow the same derivation;  $J_n$'s commute with $L_n^{(0)}$s, so those of them with $n\le 0$ can still be moved to the left where they annihilate $\< h_p|$.  Then, one commutes $z^{-q_L^2/k}$ to the left,
where it produces an extra $-z^n \frac{q_L^2}{k}$ each time it commutes with a $\partial_z$.  Descendants with $J_n$'s can always be organized so that the $J_n$ act to the right first, where it just produces a $h$-independent power of $z$ and some factors of $q_L$, so the action of the $L_n^{(0)}$s and $J_n$'s factors.  

 Thus, the conformal blocks factor into a product of Virasoro conformal blocks with $c\rightarrow c-1$ and $h_i \rightarrow h_i - \frac{q_i^2}{2k}$, and a $U(1)$ block:
 \be
 \CV_{T+J}(c, h_i, k, q_i, h_p, z) &=& \CV_T(c-1, h_i - \frac{q_i^2}{2k} , h_p, z) \CV_J(k, q_i, z), 
 \label{eq:withU1}
 \ee
 where $\CV_J$ is the contribution from just the $U(1)$ descendants $J_n$, which can be computed by directly summing the contribution at each level:
\be
\label{eq:U1Block}
\CV_J(k,q_i,z) &=& z^{-\frac{q_L^2}{k}} (1-z)^{  \frac{q_H q_L}{k}}.
\ee

We can also give a heuristic derivation of this result using background fields.  
Let us couple a classical background field $A_\mu(x)$ to a $U(1)$ current $J_\mu = \partial_\mu \phi$, where $\phi$ is a free boson.  This shifts the CFT action by $\int \phi \partial A $, leading to the equation of motion 
\be
\partial^2 \phi(x) = \partial_\mu A^\mu(x).
\ee
We obtain a classical VEV $\< J_\mu(x) \> = A_\mu(x)$.  Thus if we turn on the holomorphic $A_z(z) = - \frac{Q}{1-z}$ 
we can trivialize the 3-pt correlator $\< \CO_Q^\dag(\infty) \CO_Q(1) J(z) \>$ for an operator of charge $Q$.  This background re-absorbs the effect of $J$-exchange between $\CO_Q$ and other operators.  By writing another operator $\CO_q(x)$ of charge $q$ as $e^{i q \phi(x)} \CO_0(x)$ and evaluating the classical part of $\phi(x)$, we obtain $\CV_J$.

\section{AdS interpretation}
\label{sec:AdSInterpretation}

\subsection{Matching of backgrounds}
\label{sec:MatchingBackgrounds}

To make contact with AdS physics, it is most convenient to place the heavy operators at $z=0$ and $\infty$ in the CFT, corresponding to a heavy AdS object that exists for all time. The conformal transformation $w = z^\alpha$ from section \ref{sec:BackgroundMetric} 
has a simple geometric interpretation in AdS$_3$.  A general vacuum metric can be written as
 \be
 ds^2 = \frac{L(z)}{2} dz^2 + \frac{\bar{L}(\bar{z})}{2} d\bar{z}^2 + \left( \frac{1}{y^2} + \frac{y^2}{4} L(z)\bar{L}(\bar{z})\right) dz d\bar{z} + \frac{dy^2}{y^2}.
 \ee
A point mass source with dimension $h_H+\bar{h}_H$ and spin $h_H - \bar{h}_H$ corresponds to
\be
L(z) = \frac{1}{2z^2} \left( 1-24\frac{h_H}{c}\right) = \frac{\alpha_H^2}{2 z^2} , \qquad \bar{L}(\bar{z}) = \frac{1}{2\bar{z}^2} \left( 1-24\frac{\bar{h}_H}{c}\right)= \frac{\bar{\alpha}_H^2}{2 \bar{z}^2}.
\ee
This can be mapped to global coordinates by  $y =2 \sqrt{\frac{ z \bar{z}}{\alpha_H \bar{\alpha}_H}} e^\kappa $: 
  \be
  ds^2 &=&  d \kappa^2 +\alpha_H\bar{\alpha}_H \frac{\cosh(2 \kappa)}{2}  \frac{d z d\bar{z}}{z \bar{z}}+\frac{1}{4} \left( \alpha_H^2 \frac{dz^2}{z^2} + \bar{\alpha}_H^2 \frac{d \bar{z}^2}{\bar{z}^2} \right),
\ee
which makes it manifest that the metric becomes locally pure AdS$_3$ under $z= w^{\frac{1}{\alpha_H}}, \bar{z} = \bar{w}^{\frac{1}{\bar{\alpha}_H}}$.  Furthermore, the boundary stress tensor is related to the metric in these coordinates by $T(z) = \frac{1}{2 z^2}- \frac{c}{12} L(z)$.  The AdS interpretation of the coordinate transformation  
is therefore particularly simple: the heavy operators create a background value for $T(z)$, $\<h_H| T(z) | h_H\> = \frac{h_H}{z^2}$, which is removed when we transform to $w$ coordinates.  It is remarkable that this connection between the bulk and boundary descriptions is exhibited at the level of the individual conformal blocks.

To exhibit the black hole horizon, one can bring the metric into the original BTZ coordinates, 
\be 
 z &=& e^{t+ \phi}, \qquad 
  \bar{z} = e^{t-\phi}, \nn\\
  \kappa &=& - \frac{1}{2} \cosh^{-1} \left( \frac{4 r^2 + \alpha_H^2 + \bar{\alpha}_H^2}{2 \alpha_H \bar{\alpha}_H} \right),
  \ee
  with metric
  \be
  ds^2 &=& -\frac{(r^2- r_+^2)(r^2- r_-^2)}{r^2} dt^2 + \frac{r^2}{(r^2-r_+^2)(r^2-r_-^2)} dr^2 + r^2 \left(d\phi + \frac{r_+ r_-}{r^2} dt \right)^2.
  \ee
The inner and outer horizons $r_-, r_+$ are related to $\alpha_H, \bar{\alpha}_H$ by
\be
(r_+ +r_-)^2  &=&- \alpha_H^2  = \left( \frac{24 h}{c} -1 \right), \nn\\
(r_+- r_-)^2 &=& - \bar{\alpha}_H^2  = \left( \frac{24 \bar{h}}{c} - 1 \right).
\ee
One sees by inspection that the inner and outer horizons $r_-, r_+$ appear in the above metric only in the combinations $r_+^2 + r_-^2$ and $r_+ r_-$, which are real-valued for any $h, \bar{h}$.  However, the values $r_\pm$ themselves can be complex:
\be
r_\pm &=& \frac{\left( \frac{24 h}{c} -1 \right)^{1/2} \pm \left( \frac{24 \bar{h}}{c} -1\right)^{1/2} }{2}.
\ee
Therefore, the black hole horizon exists only when both $h$ and $\bar{h}$ are greater than $\frac{c}{24}$; in other words, to make a black hole in AdS$_3$, states must have twist $\tau$ greater than $\frac{c}{12}$.

  \subsection{Extension to $U(1)$}

The extension of the conformal blocks to include a $U(1)$ conserved current in (\ref{eq:withU1}) also has a direct connection to backgrounds in AdS$_3$.  To include a $U(1)$ current in the boundary, we need to add a gauge field in the bulk, and in order to correctly match the boundary description it must be a Chern-Simons gauge field without a Maxwell kinetic term $F_{\mu\nu}F^{\mu\nu}$.  

  Perhaps the most concrete way of understanding this is by considering the physical modes in the spectrum arising from the bulk gauge field.  A conserved spin-1 current on the boundary corresponds to a pure boundary mode in AdS$_3$, much like gravitons are pure boundary modes in AdS$_3$.  This can be seen from the fact that the wavefunction $\Phi^\mu \equiv \< 0 | A^\mu | J\>$ must  have frequency 1 and spin 1, and be annihilated by the special conformal generators.  The former condition implies $\Phi^\mu = e^{i(t\pm \theta)} f^\mu(\rho)$, and the latter condition implies \cite{Katz}
  \be
  \Phi^\mu &=& (\Phi^t, \Phi^\rho, \Phi^\theta) \propto -i e^{i(t\pm \theta)} \cos \rho \tan \rho (1 , -i \cot \rho, \pm 1)^\mu    = - \partial^\mu(e^{i (t \pm \theta)} \sin \rho).
   \ee
   This is manifestly pure gauge in the bulk, and so it is a boundary mode.  A Chern-Simons gauge field in the bulk is topological and thus has only boundary modes, as required, whereas a Maxwell gauge field has a bulk degree of freedom.  
   
   This becomes even more explicit when we consider a Chern-Simons term together with a Maxwell kinetic term in the bulk.\footnote{The Chern-Simons term is necessarily present due to the chiral anomaly, which is given by the level $k$ of the $U(1)$ boundary current and thus must be a positive integer \cite{Jensen}.  }  In this case, the bulk photon gets a topological mass, and thus the dimension of the dual CFT mode is raised above 1, implying that it cannot be a conserved current.

The calculation of correlators arising from exchange of a bulk $U(1)$ Chern-Simons fields was performed in \cite{Keranen}.  Here we will discuss the generalization to include gravity as well.  The action is
\be
S_E &=& S_{\rm bulk}+ S_{\rm bd}, \\
S_{\rm bulk} &=&  -\frac{k}{8 \pi} \int d^3 x \epsilon^{\mu\nu \lambda} (A_\mu \partial_\nu A_\lambda - \bar{A}_\mu \partial_\nu \bar{A}_\lambda) , \nn\\
S_{\rm bd} &=&  - \frac{k}{8 \pi} \int d^2 x \sqrt{\gamma}(A^2 + \bar{A}^2 - 2 A\cdot \bar{A} - 2i \epsilon^{ij} A_i \bar{A}_j) = -\frac{k}{8 \pi} \int d^2 z (A_z A_{\bar{z}} + \bar{A}_z \bar{A}_{\bar{z}} - 2 A_{\bar{z}} \bar{A}_z). \nn
\ee
The bulk term does not depend on the metric and thus the computation of the four-point function from Chern-Simons exchange in \cite{Keranen} goes through almost, but not quite, unchanged.  The result there exactly matches (up to a difference in normalization of $q_i$) the four-point function (\ref{eq:U1Block}) from summing over $J_{-n}$ descendants .\footnote{It is stated in \cite{Keranen} that the derivation is valid only at leading order in $1/k$; however, in the semi-classical limit $k\rightarrow \infty$ with $q_H/k$ and $q_L$ fixed, the backreaction from the light $\phi_L$ field on the Chern-Simons background can be neglected, and the dependence on $k$ to all orders is captured.} However, the boundary terms above do depend on the metric, and thus they feel the presence of gravity.  Their effect is very simple: in a normalization where $J_z = \frac{k}{(2\pi)^{1/2}}A_z $, they shift the boundary stress tensor by \cite{Jensen}
\be
T_{zz} &\rightarrow& T_{zz} + \frac{\delta  S_{\rm bd}}{\delta \gamma^{zz}} = T_{zz} - \frac{1}{2k} J_z J_z.
\ee
Thus, the charge $q$ of a source reduces its energy as measured at the boundary by $q^2/2k$.  This exactly matches the shift in the conformal blocks in equation (\ref{eq:withU1}).  It also implies that the matching to the spectrum in a black hole background generalizes immediately.  Because the Chern-Simons field and the metric are not coupled in the bulk, black hole solutions are just BTZ  with a shifted mass parameter, $M\rightarrow M - q^2/2k$, and a background profile for the Chern-Simons field \cite{Keranen}.  For a light neutral CFT operator in the background of a heavy charged state, this shift is the only change as compared to a heavy neutral state.  If the light operator is charged, the background Chern-Simons field produces the additional dependence (\ref{eq:U1Block}), which completely factorizes.  The spectrum of twists (or quasi-normal modes) at large spin can then be read off from the bootstrap equation exactly as in \cite{Fitzpatrick:2014vua}.  The only two changes are that the ``binding'' energy depends on $h_i - q_i^2/2k$ rather than $h_i$, and there is an additional contribution $q_B q_S/k$:
\be
\tau_n = 2 h_B+ \frac{q_Bq_S}{k} + 2i \sqrt{24 \frac{h_B - q_B^2}{2k c}-1} (n + h_S - \frac{q_S^2}{2k}) .
\ee

\subsection{Matching of correlators}

It would be interesting to see how the large $c$ Virasoro blocks arise as contributions to heavy-light 4-pt correlators obtained from Feynman diagram computations in AdS$_3$.  We previously analyzed the case of the identity block \cite{Fitzpatrick:2014vua}, but matching with more general correlators is significantly more challenging.  Here we will confine our analysis to the case of 3-pt correlators.  This will be sufficient to show that the somewhat mysterious rescaling of $\delta_H \to \delta_H/ \alpha$ in the Virasoro conformal block is a kinematical consequence of the conformal transformation to the $w$ coordinates.

In a flat background metric, the CFT 3-pt correlator is
\be
\< \CO_{H_1}(\infty) \CO_{H_2}(0)  \CO_{L}(z) \> = z^{-h_L + \delta_H} .
\ee
It is non-trivial to obtain this from an explicit bulk computation that includes gravitational backreaction, because the heavy operators have bulk-to-boundary propagators that are dressed by gravitons, and the back-reaction of the geometry on the bulk-to-boundary propagator is not suppressed.  However, conformal symmetry guarantees the form of the final result.   The conformal transformation to $w = z^\alpha$ produces a new correlator
\be
\< \CO_{H_1}(\infty) \CO_{H_2}(0) \CO_L(w) \> =w^{-h_L+\fr{\de_H}{\alpha}} .
\label{eq:3ptwCoordinates}
\ee
Let us recall how this rescaling of $\delta_H$ from $z$ to $w$ coordinates can be derived from a bulk analysis.

We will write AdS in global coordinates
\be
ds^2 = \frac{1}{\cos^2 \rho} (dt^2 + d \rho^2 + \sin^2 \rho \, d \theta^2)
\ee
where $\rho = \pi/2$ corresponds to the boundary of AdS.
To study CFT correlators in a flat metric, we approach the boundary with $\rho(\epsilon) = \pi/2 - e^{-t} \epsilon$ and then send $\epsilon \to 0$.  This produces a boundary metric
\be
ds^2 = e^{2t} (dt^2 + d \theta^2) = dz d \bar z.
\ee
The CFT operator $\CO_L$ is defined through the boundary limit of a bulk field $\phi_L(t, \theta, \rho)$ as
\be
\CO_L(t, \theta) \equiv \frac{\phi_L(t, \theta, \pi/2 - \epsilon)}{\epsilon^{\Delta}}.
\ee
The CFT correlators are obtained by applying this limit to the correlators of bulk fields.

By transforming to $w$ coordinates on the boundary, we move the CFT to a new background with metric
\be
ds^2 = dw d \bar z = \frac{dw}{dz} dz d \bar z.
\ee
To obtain the new metric from AdS/CFT, we must approach the boundary via
\be
\rho(\epsilon) = \frac{\pi}{2} - \epsilon \, e^{-t} \sqrt{\frac{dz}{dw}}
\ee
where we take the limit $\epsilon \to 0$.  
Thus we recover the conformal transformation of a primary operator
\be
\CO_L(w) = \left[ \frac{dz}{dw} \right]^{h_L}  \CO_L(z(w)) 
\ee
which gives  
\be
\< \CO_{H_1}(\infty) \CO_{H_2}(0)  \CO_{L}(w) \> \propto \left[w^{\frac{1- \alpha}{\alpha}} \right]^{h_L} w^{\frac{-h_L + \delta_H}{\alpha}}
= w^{-h_L + \frac{\delta_H}{\alpha}} 
\ee
as claimed.
This explains the rescaling of $\delta_H \to \delta_H/ \alpha$ in the Virasoro conformal block of equation (\ref{eq:SCblockFinal}) as the simple consequence of a conformal transformation.

 \section{Implications for thermodynamics}
 \label{sec:ImplicationsThermodynamics}
 
At first glance, our results suggest that in any CFT$_2$ at very large $c$, heavy operators with $h_H, \bar h_H > \frac{c}{24}$ produce a thermal background for light correlators, because in the $w = 1 -(1 -z)^{2 \pi i T_H}$ coordinates we have an explicit periodicity in the Euclidean time $t \sim \log (1- z)$.  If the full correlator had this periodicity, it would imply a version of the eigenstate thermalization hypothesis, or alternatively a derivation of BTZ black hole thermality, as reviewed in appendix \ref{app:PeriodicityinTime}.  This interpretation is too naive, but let us first note what it does immediately imply, and then move on to a more general discussion.

\subsection{Far from the black hole: the lightcone OPE limit}

As discussed in \cite{Fitzpatrick:2012yx, KomargodskiZhiboedov, Fitzpatrick:2014vua}, there is a lightcone OPE limit where correlation functions are dominated by the exchange of the identity Virasoro block.\footnote{More generally, the lightcone OPE limit is dominated by the exchange of twist-zero operators.}  In a CFT$_2$ without any conserved currents besides the stress tensor and $U(1)$s, this means that correlators such as
\be
 \< \CO_{H_1}(0) \CO_{H_2}(\infty) \CO_{L_1}(1) \CO_{L_2}(z) \> 
\ee
are thermal in the limit where $z \to 1$ with $\bar z$ fixed, or vice versa, at large central charge.  These correlators capture physical setups in AdS where the light probe is very far from the black hole.  This strongly suggests that if a thermal interpretation is appropriate, the temperature must be the $T_H$ given in equation (\ref{eq:HawkingTemperature}).

One can argue via the operator product expansion that more general correlators will have the same property in an appropriate limit.  For example, consider a 6-pt function
\be
 \< \CO_{H_1}(0) \CO_{H_2}(\infty) \CO_{L_1}(z_1) \CO_{L_2}(z_2) \CO_{L_3}(z_3) \CO_{L_4}(z_4) \>. 
\ee
In the limit that the $z_i$ approach each other in a lightcone limit,  
this correlator will be dominated by identity exchange between the heavy and light operators.  By expanding the light operators in the OPE we can reduce the 6-pt correlator to a 4-pt correlator, and conclude that the result must be thermal. A more physical way to obtain the same answer is to transform to $w$-coordinates as in section \ref{sec:BackgroundMetric}. %
The action of the Virasoro generators $\CL_n$ with $|n| \geq 2$ are trivialized by this transformation, and so 
expanding in the lightcone OPE limit we simply have 
\be
& &  \< \CO_{H_1}(0) \CO_{H_2}(\infty)  \CO_{L_1}(w_1) \CO_{L_2}(w_2) \CO_{L_3}(w_3) \CO_{L_4}(w_4) \>   \nn
\\
& & \ \ \ \ \ \ \ \ \ \ \approx  \< \CO_{H_1}(0) \CO_{H_2}(\infty) \>  \times \< \CO_{L_1}(w_1) \CO_{L_2}(w_2) \CO_{L_3}(w_3) \CO_{L_4}(w_4) \> \label{eq:6pt}
\ee
in the $w$-coordinate background.  The heavy operator produces a thermal background for the four light operators.  This can obviously be generalized to $2+n$-point correlators with two heavy operators.

\subsection{Implications for entanglement entropy}

CFT entanglement entropies can be computed via the replica trick
\be
S_A = \lim_{n \to 1} \frac{1}{n-1} \log \left( \mathrm{tr} \, \rho_A^n \right)
\ee
where $\rho_A$ is the reduced density matrix for a geometric region $A$, and we analytically continue the integer $n$.  This trick is useful because the Renyi entropies $\rho_A^n$ for integer $n$ can be computed by gluing together multiple copies of the system along the region $A$.  In CFT this means that they can be obtained from correlation functions of twist operators, which admit a conformal block decomposition.  See \cite{Calabrese:2009qy} for a review and \cite{HartmanExcitedStates,TakayanagiExcitedStates} for a relevant recent application using the Virasoro blocks.  The twist operators have fixed dimension
\be
h_n = \frac{c}{24} \left( n - \frac{1}{n} \right)
\ee
and so at large $c$, for $n \neq 1$ their dimensions are of order $c$.  As discussed in section \ref{sec:OrderofLimits}, the limit $c \to \infty$ with $h$ fixed followed by $h \to 0$ should commute with the limit where $h/c$ is fixed, followed by $h/c \to 0$.  Thus our results should apply to the computation of entanglement entropies using twist operators, and the recursion relations of section \ref{sec:RecursionFormula} could be used to obtain corrections in $1/c$.

Our result for the heavy-light Virasoro blocks are more precise than those obtained in \cite{Fitzpatrick:2014vua}, but in fact we have obtained the same formula for the identity or vacuum Virasoro block
\be
\CV_1(z) = \frac{1}{(1-z^{\alpha_H})^{2 h_L}} + \CO(1/c).
\ee
Previously the uncertainty was up to an order one multiplicative function, but it has been reduced to an additive term suppressed by $1/c$.  This implies that the corrections to \cite{Fitzpatrick:2014vua} are smaller than might have been imagined, particularly in theories with a gap in the spectrum above the identity operator.

One might also study the entanglement entropy of several intervals in the presence of a non-trivial background.  In the limits where the analysis of \cite{HartmanLargeC, HartmanExcitedStates} remains valid, so that the entanglement entropy is dominated purely by exchange of operators that are Virasoro descendants of the vacuum, one could study kinematic limits where the factorization of equation (\ref{eq:6pt}) obtains.  Thus we find that the entanglement entropy for a region $A$ composed of many disjoint intervals is \cite{HartmanLargeC} 
\be
S_A = \min \frac{c}{6}  \sum_{(i,j)}  \log \left( (w_i - w_j)^2\frac{z_i z_j}{w_i w_j} \right) 
\ee
(up to a UV-regulator-dependent normalization) where $i$ denote the end points of various intervals in the CFT, we minimize over all possible pairings $(i,j)$.
 The argument of the logarithm has the form of a distance in the $w$-metric multiplied by the Jacobian factor associated with the transformation rule of primary operators, as in equation (\ref{eq:PrimaryTransformation}).  

As one would expect from the discussion of section \ref{sec:MatchingBackgrounds}, this is simply the minimum of the sums of geodesics connecting interval end-points in the appropriate deficit angle or BTZ background.  The choice of branch cut in the logarithm determines how many times the geodesic winds \cite{HartmanExcitedStates} around the singularity.

\subsection{General kinematics and a connection with AdS locality}
 
To understand thermality for general kinematics we need to study the full correlator, which is an infinite sum over Virasoro blocks.  The difficulty is that the hypergeometric function in equation (\ref{eq:SCblockFinal}) has branch cuts in the $w$-coordinate, so periodicity in Euclidean time is not obvious for general blocks.  Furthermore, the sum over the Virasoro blocks does not converge everywhere in the $z, \bar z$ planes, so the sum does not necessarily have the same properties as its summands.  Let us study these issues in more detail, and then we will formulate some general, sufficient criteria for thermality, motivated in part by our present understanding \cite{Heemskerk:2009pn, Heemskerk:2010ty, ElShowk:2011ag, JoaoMellin, Sundrum:2011ic, analyticity, AdSfromCFT} of the relationship between CFT data and AdS locality.

 The full Virasoro blocks are products $\CV_h (z) \CV_{\bar h} (\bar z)$, and the Euclidean time coordinate is 
 \be
 t = \log((1-z)(1- \bar z )) = -\frac{i}{2 \pi T_H} \log((1-w)(1- \bar w ))
 \ee  
where we are considering the scalar case $h_H = \bar h_H$ and $\delta_H = 0$ for simplicity.  Translation in Euclidean time corresponds to rotation about point $w, \bar w = 1$.   To study the branch cut structure, it is useful to apply the hypergeometric identity\footnote{There is a logarithm in the special case where $\gamma - \alpha - \beta$ is an integer.}
  \be
  {}_2F_1(\alpha, \beta,\gamma,1-x) &=& \frac{\Gamma(\gamma)\Gamma(\gamma-\alpha-\beta)}{\Gamma(\gamma-\alpha)\Gamma(\gamma-\beta)} {}_2F_1(\alpha, \beta, \alpha+\beta-\gamma+1,x) \\
  && + \frac{\Gamma(\gamma)\Gamma(\alpha+\beta-\gamma)}{\Gamma(\alpha)\Gamma(\beta)} x^{\gamma-\alpha-\beta} {}_2F_1(\gamma-\alpha, \gamma-\beta, \gamma-\alpha-\beta+1,x) \nn
  \ee
to the Virasoro blocks. The hypergeometric functions on the right hand side have a convergent Taylor expansion around $x=0$, so the only branch cut emanating from $x = 0$ arises from the explicit power $x^{\gamma-\alpha-\beta}$.  Applying this relation to $\CV_h (w) \CV_{\bar h} (\bar w)$ produces four terms, which can be written schematically as
\be \label{eq:ExplicitBlockMonodromy}
\CV_h (w) \CV_{\bar h} (\bar w) &=& A(w) \bar A(\bar w) + (1-w)^{2 \delta_L} (1 - \bar w)^{2 \bar \delta_L} B(w) \bar B(\bar w)
\nn \\ &&
+ (1-w)^{2 \delta_L} \bar B(w)  A(\bar w) + (1 - \bar w)^{2 \bar \delta_L} A(w) \bar B (\bar w)
\ee
where the functions $A, \bar A, B, \bar B$ have trivial monodromy about $w, \bar w=1$.
The quantity $2 \delta_L - 2 \bar \delta_L$ must be an integer if the light operators have half-integer or integer spin, so the terms on the first line are automatically periodic under $ t \to t + 1/T_H$.  The terms on the second line are problematic due to the branch cut emanating from $w, \bar w = 0$.  These terms must cancel in the sum over Virasoro blocks in order to give a thermal correlator.

It should be emphasized that these branch cuts are a general issue for any CFT correlator written as a sum over conformal blocks.  The Virasoro blocks at large $c$ are basically identical to global conformal blocks written in a new coordinate system.   
Correlators of integer spin operators must be periodic in the angle $\theta = \log(1-z) - \log(1- \bar z)$, and the terms on the second line of equation (\ref{eq:ExplicitBlockMonodromy}) do not have this property when we replace $w \to z$ to turn the Virasoro blocks back into global blocks.  In conventional CFT correlators, the branch cuts from these problematic terms must always cancel to restore periodicity in $\theta$.  We would like to know when they cancel in the heavy-light correlators, since this would imply thermality.

In theories with a sparse low-dimension spectrum, one would expect that few conformal blocks contribute, and so the sum over all the blocks has simple properties.  There has been major progress  understanding such theories using the CFT bootstrap, and in particular, it has been shown  that the offending branch cuts in the analogue to equation (\ref{eq:ExplicitBlockMonodromy}) for the global conformal blocks in $z$ coordinates are canceled by infinite sums over `double-trace' operators.  This phenomenon can be generalized far beyond the class of theories with a sparse spectrum by using Mellin amplitudes \cite{Mack, Macksummary, JoaoMellin, NaturalLanguage, Paulos:2011ie} to represent CFT correlators, as we will discuss in a forthcoming paper \cite{Eikonalization}.   

Any CFT whose correlators have exponentially bounded Mellin amplitudes will be guaranteed to have physical periodicity in $\theta$.
 As a toy example, consider the function $f(z)$ 
\be
f(z) = \int_{-i \infty}^{i \infty} d s \, M(s) (1-z)^{-s}
\ee
written in terms of a Mellin transform.  The function $f$ will not have any branch cuts emanating from $z=0$ if $M(s)$ falls off exponentially fast as $s \to \pm i \infty$.  In other words, the asymptotic behavior of the Mellin transform $M(s)$ determines the analyticity of the original function.   This fact will be used \cite{Eikonalization} in forthcoming work to study universal properties of CFT correlators and OPE coefficients.  

The asymptotic behavior of the Mellin amplitude also has a close connection with short-distance AdS locality \cite{analyticity, AdSfromCFT}.  CFTs with a perturbative expansion parameter, a Fock space spectrum, and a polynomially bounded Mellin amplitude can always be described by an effective field theory in AdS.  Exponential boundedness is a weaker condition that nevertheless implies the desired periodicity in $\theta$.  

Given the simple relationship between Virasoro and global blocks, including the equivalent issues with periodicity in $t$ and $\theta$, it is natural to consider a Mellin amplitude representation directly in the $w$ coordinates. This is also motivated by the idea that the heavy operator simply creates a new classical background in which the dynamics transpire. In analogy with the conformal cross ratios $u= z \bar z$ and $v = (1-z)(1- \bar z)$, we write the heavy-light correlator
\be \label{eq:wMellin}
 \< \CO_{H_1}(0) \CO_{H_2}(\infty) \CO_{L_1}(1) \CO_{L_2}(w, \bar w) \> = \int_{-i \infty}^{i \infty}  ds dt \, M_w(s, t) [w \bar w]^{-s} [(1-w)(1-\bar w)]^{-t}
\ee
using the Mellin space representation $M_w(s, t)$.  As long as $M_w(s,t)$ falls off exponentially as $s, t \to \pm i \infty$, the correlator will be analytic around $w, \bar w = 0$.  Note that periodicity in $\theta$ is not obvious in this representation, but imposes further constraints on $M_w$.  We stress that $M_w$ is not the usual Mellin amplitude $M$: the two differ by whether the Mellin transform is performed on the correlator in $z$ coordinates or in $w$ coordinates, and exponential boundedness of one does not necessarily translate into exponential boundedness of the other.

When written in $w, \bar w$ coordinates as in equation (\ref{eq:wMellin}), the Virasoro blocks have a simple Mellin amplitude representation $B_w(s,t)$ \cite{Mack, analyticity}.  The existence of specific operators in the light-light and heavy-heavy OPE limits implies the presence of specific poles in $B_w(s,t)$, and these poles must remain in the Mellin representation of the full CFT correlator.  The individual $B_w(s,t)$ do not fall off exponentially as $s, t \to \pm i \infty$, since the Virasoro blocks are not analytic near $w = 1$.  But if the sum of the $B_w(s,t)$ has a good asymptotic behavior, then the full CFT correlator will be thermal.  One might also investigate factorizations as in equation (\ref{eq:6pt}) in this Mellin space language.

\section{Future directions}
\label{sec:FutureDirections}

Intuition suggests that CFT operators carrying a large conserved charge may generate a classical background field affecting correlation functions in a universal way.  This idea has obvious validity in the case of AdS/CFT, where heavy or charged states generate gravitational or gauge fields in AdS, but one can also look for manifestations of this effect in general CFTs.  We have shown that heavy operators in 2d CFTs have exactly this effect, and that it can be used to compute Virasoro conformal blocks at large central charge.  We also used this idea to sum up contributions from $U(1)$ currents in the CFT.  Classical gravitational and gauge fields in AdS$_3$ have a universal manifestation in all CFT$_2$ at large central charge.

To leading order at large central charge $c$, the heavy-light Virasoro blocks bear a striking imprint of thermality, and it would be interesting to explore this property in more detail.  In particular, one should expect that the probability of a transition from one heavy state to another should be weighted by a Boltzmann factor.  The rescaled
$\frac{\delta_H}{\alpha} = 2 \pi i \frac{E_{H_1} - E_{H_2}}{T_H}$ 
appearing in the conformal blocks of equation (\ref{eq:SCblockFinal}) seems tantalizing in this regard.  
In fact, in the limit of small $h, \delta_L$ the conformal blocks approach a universal form
proportional to $(1-w)^{\delta_H/\alpha}$.  This result might be used to argue that the rates for transitions among the heavy states take a universal thermal form.

Unitary CFT correlators should not be exactly thermal.  One can use the recursion relations of section \ref{sec:RecursionFormula} to compute corrections to our $c=\infty$ result, but at present this is only practical if we work to low order in the coordinate $w$, or perhaps to first order in $1/c$.  However, corrections to thermality are most likely non-perturbative in $c$, showing up in the Virasoro blocks at finite $z$ or $w$.  It would be very exciting to understand these non-perturbative corrections.  CFT correlators are made from an infinite sum of Virasoro blocks, so it would also be interesting to understand whether corrections to thermality arise primarily from non-perturbative corrections to the individual Virasoro blocks, or from patterns in the infinite sum.

An immediate question is whether one can obtain similar results using classical background fields for CFTs in $d>2$ dimensions.  We were able to obtain very precise results for CFT$_2$ by making use of the infinite dimensional Virasoro algebra, so one might not expect higher dimensional generalizations to be possible.  However, we believe that the results of \cite{Brown:1977sj, Herzog:2013ed}    on the expectation value of the stress-tensor in conformally flat backgrounds can be used to study heavy operators in large central charge CFTs in $d=4$ and $6$, perhaps after imposing some physical constraints \cite{Camanho:2014apa} on the CFT data.  It would be very interesting to generalize the analysis of \cite{Brown:1977sj, Herzog:2013ed, Huang:2013lhw}, and to understand which CFT data affect the result.

Our analysis motivates studying CFTs in a non-trivial background metric chosen to trivialize stress tensor correlators.  We have worked with a pure CFT, but one might also consider a CFT with a UV cutoff, or a general local QFT.  By recapitulating the computations of section \ref{sec:BackgroundMetric} in a CFT with a UV cutoff, one might attempt to derive the bulk Einstein equations.  The idea would be to identify the cutoff with the position of a `UV brane', and determine an appropriate metric as a function of the cutoff.  Perhaps one could obtain a more precise understanding of the relationship between UV cutoff schemes and bulk locality. 

There is also a fascinating connection between conformal blocks and entanglement entropy at large central charge \cite{Headrick} which has recently led to increasingly rich matching of such effects between gravity and field theory computations \cite{HartmanLargeC, HartmanExcitedStates,TakayanagiExcitedStates, DongGravityRenyi, Chen:2013kpa, KrausBlocks}.   We have discussed how an extension to entanglement entropies of multiple intervals in the background of an excited state follows naturally from the results here, and it would be interesting to consider connections to other deformations as well.  For instance, the addition of light states in the gravity theory produces quantum corrections to the Ryu-Takayanagi formula \cite{XiYinGravCorrections1,XiYinGravCorrections2,JuanGravCorrections}, and in terms of the field theory these should involve conformal blocks for states with corresponding conformal weights; it would be interesting to investigate the precise decomposition.  Finally, even within the pure gravitational theory, loop corrections to Renyi entropies on the gravity side are reproduced by conformal blocks including not just Virasoro descendants of the vacuum but also additional descendants that arise from acting with operators from the different ``replicas'' \cite{DongGravityRenyi, Chen:2013kpa}.  It seems natural to think of these contributions as descendants of the vacuum under an enhanced symmetry, and to try to simplify the field theory computation using some of the methods employed here.

It would be interesting to study more general background field configurations, and couplings that source more general CFT operators. One can also use the methods of section \ref{sec:BackgroundMetric} to find more complicated background metrics that simplify stress tensor correlators in the presence of more than two heavy operators, for instance by combining global conformal transformations with repeated application of $z \to z^\alpha$.  Such conformal mappings can be used to study states created by several heavy operators.  Supersymmetric theories and higher spin conserved currents would also be of interest, and our methods could provide interesting corrections to the results of  \cite{deBoer:2014sna} on conformal blocks in theories with $\mathcal{W}_N$ symmetry.

\section*{Acknowledgments}

We would like to thank Chris Brust, Liang Dai, Tom Faulkner, Shamit Kachru, Ami Katz, Zuhair Khandker, Logan Maingi, Gustavo Marques Tavares, Junpu Wang, and Xi Dong for valuable discussions.  JK is supported in part by NSF grant  PHY-1316665 and by a Sloan Foundation fellowship.  ALF was partially supported by ERC grant BSMOXFORD no. 228169. MTW was supported by NSF grant PHY-1214000 and DOE grant DE-SC0010025.

\appendix 
 
 \section{Details of recursion formula} 
 \label{app:recursion}

In this appendix we will present explicit formulas necessary for the evaluation of the recursion formula, which were first derived in \cite{ZamolodchikovRecursion,Zamolodchikov:1987}. Instead of using the central charge $c$, it is often more convenient to work with the parameter $T$:
\be
c = 13 - 6(T+1/T).
\ee
For each $c$, there are clearly two solutions for $T$, and which one we use will be defined in context momentarily.

The poles in $c$ arise from degeneracy conditions at special values determined by Kac.  For any positive integers $m$ and $n$, there is a degenerate state when
\be
h_p &=& \frac{c-1}{24} + \frac{(\alpha_+ m + \alpha_- n)^2}{4},
\ee
where 
\be
\alpha_\pm &=& \frac{1}{\sqrt{24}}\left( \sqrt{1-c} \pm \sqrt{25-c}\right).
\ee
Note that $\alpha_+ \alpha_- = -1$.  Now, we will define $T$ to be the solution such that
\be
\alpha_+ = \sqrt{T}, \qquad \alpha_- = -1/\sqrt{T}.
\ee
Thus, the degeneracy condition can be written
\be
h_p &=& - \frac{(1-T)^2}{4T} + \frac{(m \sqrt{T} - n /\sqrt{T})^2}{4}.
\label{eq:deg}
\ee
This has the solution
\be
T_{m,n}(h_p) &=& \frac{2h_p +mn-1 + \sqrt{(2h_p +mn-1)^2 - (m^2-1)(n^2-1)}}{n^2-1}.
\label{eq:TmnPlusBranch}
\ee
The poles values $c_{m,n}(h_p)$ are then just given by
\be
c_{m,n} &=& 13 - 6(T_{m,n}(h_p)+1/T_{m,n}(h_p)).
\ee
There is another branch of solutions to (\ref{eq:deg}) corresponding to a plus sign in front of the square root. However, it is straightforward to verify that exchanging branches is equivalent to 
%this corresponds to the same value of $c$ as if we
 interchanging $m$ and $n$.  Thus, if we consider all pairs of integers $m,n$ such $m\ge 1$ and $n\ge 2$, then we cover all the degeneracy values $c_{m,n}(h_p)$ using just the branch in (\ref{eq:TmnPlusBranch}). 
%
%\be
%T_{m,n}(h_p) &=& \frac{2h_p +mn-1 + \sqrt{(2h_p +mn-1)^2 - (m^2-1)(n^2-1)}}{n^2-1},
%\label{eq:TmnPlusBranch}
%\ee
%for all integers $m,n$ with $m\ge 1, n\ge 2$. 
The dependence of the coefficient factors $R_{m,n}$ on the external weights is completely fixed by the fact that they must vanish at specific values allowed in theories with $c=c_{m,n}(h_p)$, together with the fact that according to the OPE, the coefficient of each power of $z$ must be a polynomial of known order in the $h_i$'s.  The remaining dependence on $T$ was determined in \cite{ZamolodchikovRecursion,Zamolodchikov:1987} to be
\be
R_{m,n}(c,h_i) &=& {\rm Jac}_{m,n} \prod_{p,q} (\lambda_1+\lambda_2 - \frac{\lambda_{pq}}{2})(\lambda_1-\lambda_2 - \frac{\lambda_{pq}}{2})(\lambda_3+\lambda_4 - \frac{\lambda_{pq}}{2})(\lambda_3+\lambda_4 - \frac{\lambda_{pq}}{2}) \nn\\
&\times &  -\frac{1}{2} \prod'_{k,\ell} \lambda_{k\ell}^{-1} .
\ee
The product over $p,q$ in the first line is from $p=-(m-1),-(m-3), -(m-5), \dots, m-3, m-1$, and $q=-(n-1),-(n-3), -(n-5), \dots , n-3, n-1$.  The primed product over $k,\ell$ in the second line is from $k=-(m-1), -(m-2), \dots, m$ and $\ell = -(n-1), -(n-2), \dots,n$, but omitting the pairs $(k,\ell)=(0,0)$ and $(k,\ell)=(m,n)$. The $\lambda_i$ and $\lambda_{p,q}$'s denote
\be
\lambda_i &=& \sqrt{h_i + \frac{(1-T)^2}{4T}}, \nn\\
\lambda_{p,q} &=& q\sqrt{T}-p/\sqrt{T}.
\ee
The `Jacobian' factor ${\rm Jac}_{m,n}$ is $\left( \frac{dc_{m,n}(h_p)}{dh_p}\right)^{-1}$:
\be
{\rm Jac}_{m,n} = \frac{dh_p}{dc} = \frac{dh_p/dT}{dc/dT}  = \frac{24(1-T_{m,n}^2)}{-1+m^2 - T_{m,n}^2(-1+n^2)}.
\ee
%The total factor is 
%\be
%R_{m,n}(c,h_i) &=& R'_{m,n}(c,h_i)/{\rm Jac}_{m,n}^{-1}(T).
%\ee

\section{Periodicity in Euclidean time and thermodynamics}

\label{app:PeriodicityinTime}

The simple thermodynamic relation
\be \label{eq:BasicThermo}
dS = \frac{dE}{T}
\ee
implies that if we know the effective temperature as a function of energy, $T(E)$, then we can relate it to $S(E)$, giving the density of states.

We will now review a more formal argument for this result based entirely on the periodicity of correlation functions in Euclidean time.
Let us assume that a correlator
\be
\< \CO_H | \CO(t) \CO(0) | \CO_H \> 
\ee
is periodic in imaginary $t$ with period $\beta$.   We will demonstrate the relation between $\beta$ and the density of states in equation (\ref{eq:BasicThermo}).

We can write the 4-pt correlation function as
\be
\< \CO_H | \CO(t) \CO(0) | \CO_H \>  &=& \int d E' e^{S(E')} \< E | \CO(t) | E' \> \< E' | \CO(0) | E \>
\\
&=& \int d E' e^{S(E')} e^{i (E - E') t} |\< E | \CO(0) | E' \> |^2 
\ee
where we wrote $\CO_H$ as the state $E$.  If we assume that the original correlator is order one, while $S(E)$ is a large number, then we expect
%\be
%\< E | \CO | E' \> \sim e^{- S(E')/2}
%\ee
%although this result is very rough.  For example, it is possible that a few correlators are order one while the others are much smaller.  To get a symmetrical answer we expect
\be
\label{eq:OffDiagonalSize}
\< E | \CO | E' \> \sim e^{- (S(E') + S(E))/4} \ \ \mathrm{or} \ \ e^{- S\left(\frac{E + E'}{2} \right)/2}
\ee
where we imposed symmetry under $E \leftrightarrow E'$, but we cannot be more precise without more information about the theory.  Of course it remains possible that most of these matrix elements are identically zero, while others are much larger than this estimate, but here we are assuming continuity in $E$ and $E'$.

The periodicity in Euclidean time, or more precisely the KMS relation, implies that
\be
\< E | \CO(t - i \beta) \CO(0) | E \> = \< E | \CO(-t) \CO(0) | E \>.
\ee
This means that we can write
\be
\int d E' e^{S(E')} e^{-i (E - E') t} |\< E | \CO(0) | E' \> |^2 = \int d E' e^{S(E')} e^{i (E - E') t + \beta (E - E')} |\< E | \CO(0) | E' \> |^2.
\ee
Now let us inverse Fourier transform from $t \to \omega$.  We obtain
\be
e^{S(E-\omega)} |\< E | \CO(0) | E - \omega \> |^2 = e^{S(E+\omega)} e^{- \beta \omega } |\< E | \CO(0) | E + \omega \> |^2.
\ee
Clearly when $\omega = 0$ this equation becomes trivial, so we do not learn anything about the overall magnitude of $\< E | \CO | E \>$ from it.
This tells us that
\be
e^{S(E-\omega) - S(E + \omega) + \beta \omega} =  \frac{ |\< E | \CO(0) | E + \omega \> |^2}{ |\< E | \CO(0) | E - \omega \> |^2}.
\ee
Now if we assume that the $\< E | \CO(0) | E' \>$ correlators take either of the forms from equation (\ref{eq:OffDiagonalSize}) and expand in $\omega$, we get the same result.  Using the first form, we obtain
\be
 e^{S(E-\omega) - S(E + \omega) + \beta \omega} =  e^{\frac{1}{2}[S(E-\omega) - S(E+\omega)]}
\ee
so that we find
\be
\beta = \frac{\partial S}{\partial E}.
\ee
%Presumably we were supposed to define our $S$ and $\beta$ so that we didn't get the extra factor of $1/4$, but otherwise this was the result we wanted.  
This result relates the periodicity in imaginary time to the density of states, identifying $\beta = 1/T$.  The additional assumptions we needed were that we can view the correlator as the integral of a smooth function of energy $E'$, with a smooth entropy function $S(E)$, and also the corresponding rough estimate of OPE coefficients from equation (\ref{eq:OffDiagonalSize}).  

\bibliographystyle{utphys}
\bibliography{LargeCBlocks}

\end{document}